# 36° step size of proton-driven c-ring rotation in $F_oF_1$-ATP synthase

Monika G. Düser[a], Nawid Zarrabi[a], Daniel J. Cipriano[b], Stefan Ernst[a], Gary D. Glick[c], Stanley D. Dunn[b] and Michael Börsch[a*]

[a] 3. Physikalisches Institut, Universität Stuttgart, Pfaffenwaldring 57, 70550 Stuttgart, Germany;

[b] Department of Biochemistry, University of Western Ontario, London, Ontario, Canada N6A 5C1;

[c] Department of Chemistry, University of Michigan, 930 N University Ave, Ann Arbor, MI 48109-1055, USA

Synthesis of the biological "currency molecule" adenosine triphosphate ATP is accomplished by $F_oF_1$-ATP synthase. In the plasma membrane of *Escherichia coli*, proton-driven rotation of a ring of 10 *c* subunits in the $F_o$ motor powers catalysis in the $F_1$ motor. While $F_1$ uses 120° stepping, $F_o$ models predict a step-by-step rotation of *c* subunits 36° at a time, which is here demonstrated by single-molecule fluorescence resonance energy transfer.

The difference of the electrochemical potential of ions across a membrane is utilized by $F_oF_1$-ATP synthase to catalyze ATP formation[1]. In current models, a flux of protons through the membrane-embedded $F_o$ part causes rotation of its ring of *c* subunits[2,3]. This rotation is mechanically coupled to the γ and the ε subunits co-rotating within $F_1$, which induce and synchronize sequential conformational changes in three catalytic binding sites resulting in the formation and release of ATP[4] (Fig. A).

Evidence for *c* subunit rotation in single $F_oF_1$ was reported previously associated with the reverse chemical reaction, i.e. ATP hydrolysis. Using video microscopy of a large fluorescent pointer, i.e. µm-long actin filaments or polystyrene double-beads, surface-attached enzymes were observed in the presence of detergent[5,6] or embedded in lipid bilayer fragments. Step sizes of 120 degrees were reported in these studies[7]. To enable ATP synthesis by ion gradients across the membrane, ion-impermeable lipid vesicles and small markers to circumvent viscous drag limitations have to be used. Accordingly, $Na^+$-driven *c* subunit rotation in single ATP synthase was monitored using the fluorescence anisotropy of a single fluorophore[8]. However, individual steps during ATP synthesis could not be determined at that time, because sampling over several individual enzymes was required to improve the signal-to-noise ratio for the anisotropy fluctuation analysis.

We have developed a single-molecule fluorescence resonance energy transfer (FRET) scheme for load-free detection of subunit rotation during ATP synthesis[9-12]. One fluorophore is attached to a rotor subunit and the second one to a stator subunit in *E. coli* $F_oF_1$-ATP synthase. Upon subunit rotation, the proximity of the fluorophores varies with time and anti-correlated changes in the relative fluorescence intensities $I_D$ and $I_A$ of FRET donor and acceptor dye are observed. Their actual distance is calculated from the time trajectory of the FRET efficiency $E_{FRET}=I_A/(I_A+I_D)$ according to the Förster theory[13]. Singly enzymes are reconstituted into liposomes, and ATP synthesis is started by mixing two buffers to initiate the pH difference plus additional electrical potential across the membrane[14]. Single-molecule FRET is measured in a confocal microscope setup, where freely diffusing liposomes generate bursts of photons while traversing the laser focus or detection volume, respectively.

To unravel the "one-after-another" stepping of proton-driven *c* ring rotation, we fused EGFP as FRET donor to the C terminus of the *a* subunit[12]. Alexa568-maleimide as FRET acceptor was covalently bound to a cysteine (E2C mutation) of one *c* subunit. ATP synthesis activity of this mutant enzyme was slightly reduced by 15% compared to the wild type enzyme (i.e. 27 ATP s$^{-1}$). Characteristic confocal FRET data from active, single $F_oF_1$-ATP synthases are shown in Fig. B and C. The large total fluorescene intensity fluctuations within the photon burst are due to free Brownian motion of the liposome. They are caused by the position dependences of excitation power and detection efficiency in the confocal volume. However, anti-correlated fluorescence changes clearly indicated intramolecular distance changes due to *c* subunit rotation during ATP synthesis, which was driven by a pH difference from pH 4.7 inside the liposome to pH 8.8 outside and in the presence of 100 µM ADP and 5 mM phosphate.

Due to the sub-stoichiometric labeling of the *c* subunit, 24% of all 11959 $F_oF_1$ contained both dyes as scrutinized by pulsed alternating lasers excitation[15,16] (see Supplementary Information). Of these, 14% of the photon bursts showed changes in FRET efficiencies. In contrast, after incubation with 60 µM DCCD only 5.2% of the bursts with donor and acceptor fluorescence showed FRET fluctuations corresponding to an inhibition efficiency of 63%. The FRET efficiency changes from step to step were smaller than found previously for the case of γ or ε subunit rotation. Sometimes up to five FRET levels occurred in a series, spanning the maximum expected range for distance changes during *c* rotation (as seen in Fig. B, C, and in Supplementary Information).

FRET level changes were assigned manually and dwell times were binned to 4 ms time intervals. The distribution is shown in Fig. D. The average dwell time was determined to 9±1 ms by fitting this histogram with one decay component. This is much shorter than the dwell times found for the γ or ε subunits in $F_oF_1$ (between 18 to 51 ms[10,11]), which rotate in 120° steps. We note that adding an additional rising component apparently improved the fitting at shorter dwells. This is due to the fact that the lower limit to determinining dwell times is 2 ms, and thus the histograms lack those data points. Assuming that one proton is transported with each *c* subunit step and that four protons have to be translocated across the membrane for each ATP[14], an ATP synthesis rate of 36 s$^{-1}$ is calculated from the single-molecule FRET data, which is in accordance with the biochemical measurements. In the presence of 20 µM of the inhibitor aurovertin B[17] (which binds to β in the $F_1$ part[18,19]) , the relative number of ATP synthases with rotating *c* subunits was reduced by 75% and the dwell times were prolonged more than 2-fold. Fitting yielded a decay component of 19±2 ms (Fig. E).

What is the step size of c ring rotation during ATP synthesis? We calculated the distance between the fluorophores on $F_oF_1$ for each FRET level and plotted FRET distance 1 against the subsequent FRET distance 2 in the FRET transition density plot[20] (Fig. F). Many FRET transitions were located near the diagonal. Two models were considered that had to be discriminated by this plot: model I expected 120° stepping for *c* as deduced from the step size of the γ and ε subunits. Model II anticipated individual step sizes of 36° as expected from the 10-fold symmetry of the *c*-ring and the assumption that each *c* subunit passes the stator individually. Geometrical constraints were a *c*-ring radius of 2.5 nm and EGFP on the cytoplasmatic side, about 4.2 nm above the plane of Alexa568 rotation, and with the chromophore centered at 3.2 nm off the rotation axis (see Supplementary Information). Given the Förster radius $R_0$ = 4.9 nm[12], this yielded the possible transitions for the case of 36° stepping shown as the white curve and 120° stepping shown as the black curve (Fig. F). The maxima in the histogram matched almost perfectly the curve for 36° steps. Some transitions were detected apart from the curve for 36° steps but this can be explained by apparent larger step sizes due to un-assigned short steps. Monte Carlo simulations for all step sizes strongly support the 36° step size and reproduced the experimental FRET transition density plot (see Supplementary Information).

In summary, our single-molecule FRET approach revealed a step-by-step rotary motion of the *c* subunit ring in 36° steps during proton-driven ATP synthesis. This result provides experimental support for the concept that, as each subunit of the decameric *c*-ring rotates past the *a* subunit, one $H^+$ is translocated, winding up the elastic element(s) of the enzyme, most likely the globular region of the γ subunit located between the *c*-ring and the $\alpha_3\beta_3$ catalytic domain[6]. We believe that we were able to detect this step-by-step motion most clearly during ATP synthesis because successive steps were slowed by the conservation of energy during this process. Further analysis may provide insight into energy storage in the system, and into how the symmetry mismatch between $\alpha_3\beta_3$ and the *c*-ring stoichiometry, present in ATP synthase from most species, affects sequential steps of ATP synthesis.

**Acknowledgements**

We thank P. Gräber and coworkers (University of Freiburg) for their help in enzyme purification and J. Wrachtrup (University of Stuttgart) for supporting the development of the confocal setup. Financial support from the Deutsche Forschungsgemeinschaft (grant BO 1891/10-1 to M.B.), the Landesstiftung Baden-Württemberg, the Canadian Institutes of Health Research (grant MT-10237 to S.D.), and the NIH (grant RO1-AI 47450 to G.D.G) is gratefully acknowledged.

**Figure legends**

Single-molecule FRET approach to detect the 36° step size of the rotary *c* subunits in $F_oF_1$-ATP synthase during ATP synthesis. **A**, model of FRET-labeled *E. coli* $F_oF_1$-ATP synthase with EGFP (green; fused to the C terminus of subunit *a*, orange) and Alexa568 (red, at residue E2C) at one of the *c* subunits (blue). Rotation of *c* results in stepwise distance changes to EGFP. **B** and **C**, photon bursts of single $F_oF_1$-ATP synthase during ATP synthesis. Lower panels show fluorescence time trajectories for EGFP (green trace) and Alexa568 (red trace), upper panels show the corresponding intramolecular distances (most-likely distance in red and deviations as blue band). **D**, dwell time distributions of FRET levels in $F_oF_1$-ATP synthases during ATP synthesis (red bars), and **E**, with 20 µm aurovertin B (gray bars, monoexponential fits in black). **F**, FRET transition density plot for proton-driven *c* subunit rotation with constraint curves for 36° steps in white and for 120° steps in black.

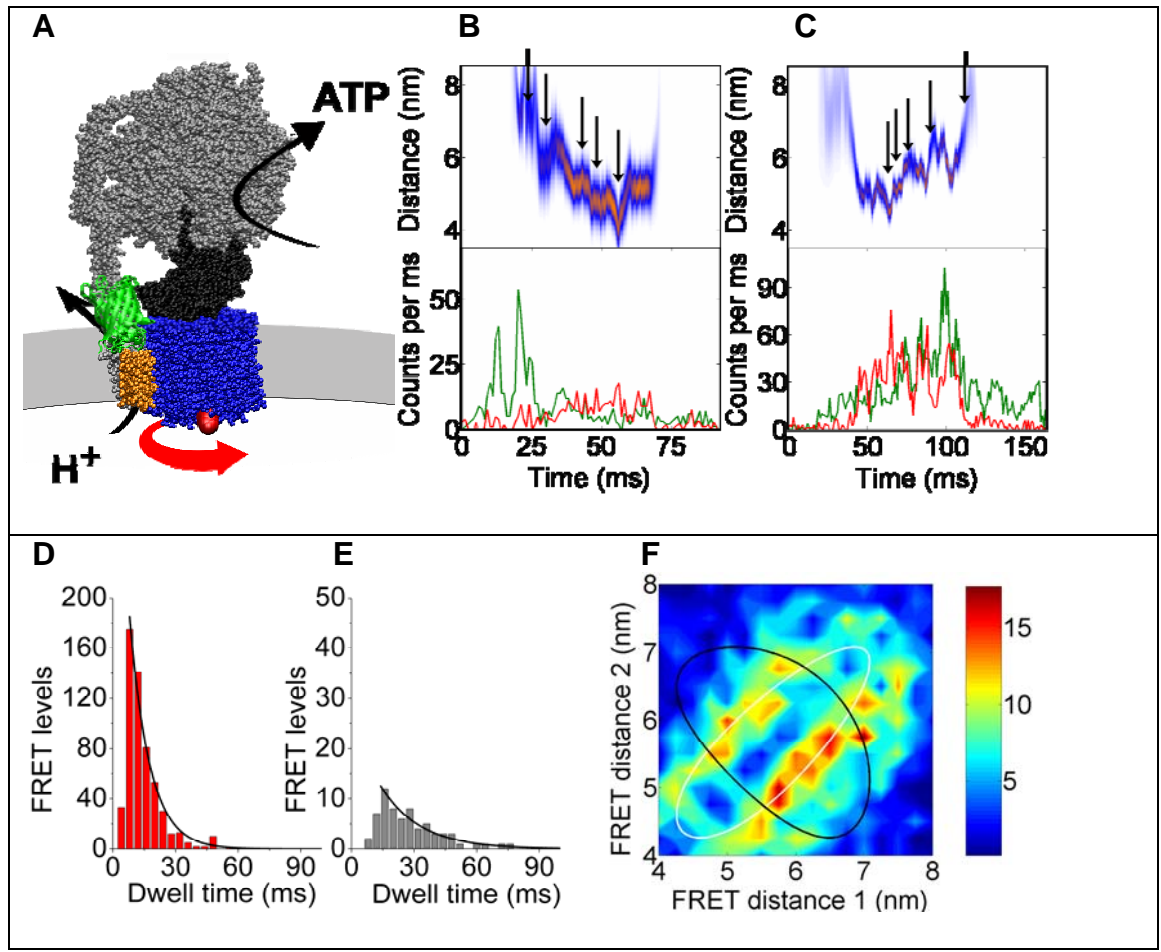